\documentclass[galley,grl]{agu2001}
\authorrunninghead{FREIRE}

\titlerunninghead{On Iapetus}

\authoraddr{P. C. C. Freire,
Arecibo Observatory, HC 03 Box 53995
Arecibo, Puerto Rico, 00612, USA
(pfreire@naic.edu)}

\begin{document}

%
%
%
%
%

%
%

\title{Solving the mystery of Iapetus}
%

%
%


\author{P. C. C. Freire}
\affil{Arecibo Observatory, Arecibo, Puerto Rico, USA}

\begin{abstract}

Since the discovery of Iapetus by G. D. Cassini, in 1672, it has been
known that the leading hemisphere of this Saturnian satellite is one order of
magnitude darker than the trailing hemisphere. Since the Cassini spacecraft
entered the Saturnian orbit, several high-quality images of the dark
hemisphere of Iapetus have been obtained, in particular during the Dec
31 2004 flyby of this satellite. These images revealed the presence of
a large equatorial ridge in the dark hemisphere of Iapetus.
We propose that this ridge and the dark
coating of the hemisphere on which it lies are intimately interlinked and are
the result of a collision with the edge of a primordial Saturnian
ring, ultimately caused by a sudden change in the orbit of
Iapetus. The model naturally explains all of the the unique features
of this satellite; it is probably the solution to one of the oldest
mysteries in solar system astronomy.

\end{abstract}

%
%

%

\begin{article}

%
%

\section{Introduction}

Recently, the dark hemisphere of Iapetus has been imaged in detail by
the Cassini spacecraft. One of the main scientific goals
of this flyby was to gather information pertinent to the solution of
the hemispheric dichotomy. Instead, to everyone's surprise, an
extraordinary ridge was found. This feature is notorious in
many ways: it follows, without deviation, a great circle along the
surface of Iapetus, and this great circle happens to be exactly
aligned with the equator. Its height is remarkable, more than 20 km in
some points. This feature has no known counterparts in the solar system.
The fact that the dark region of Iapetus (aptly named Cassini Regio)
has a perfect bilateral (North-South) symmetry relative to the
equatorial ridge suggests very strongly that the two features are
intimately linked (Porco et al. 2005).

\section{The hypothesis.}

We believe that the newly discovered ridge is the key to understanding
Cassini Regio. We suggest that both have a common origin, and that
this is a collision with a primordial ring of Saturn. In this
scenario, the ridge should, as observed, be saturated with craters,
many of these would have been produced by collisions of the ring
particles with the surface of Iapetus.

A collision with a ring is fundamentally different than a collision
with a single object, or even a string of objects caused by tidal
disruption, like Comet Shoemaker-Levy 9. In that case, we have several
relatively large objects along a single line; when these impact the
surface, they produce a carter chain. In a crater, the ejecta expands
radially from the center, which to some extent prevents accumulation
of material in the impact region (since the area where the ejecta
accumulates
is of the order $r^2$, where $r$ is the radius of the ejecta blanket).
In the case of a collision with a ring, we have much smaller craters
(because of the much smaller size of the ring particles) and the
emergence of new collective effects: when millions of craters are
being produced every second along a line, the pattern generated by the
debris will have bilateral symmetry relative to that line. Therefore,
the area where the ejecta accumulates increases much more slowly (with
$r$), leading to a much larger accumulation of ejecta per unit surface
closer to the impact line.

Another major difference is the sustained, and much larger, flux of
matter. Let us suppose a ring with a surface mass density of 1000~kg~m$^{-2}$
(equivalent to a width of 100 meters and an average spatial density 
one hundredth that of water). Let us also imagine that an object in an
equatorial orbit flies for about 10000 seconds at a speed $v$ of 2.5
km/s through such a feature. The volume of material accumulated per
meter along the equator would be of the order of $2.5 \times 10^{7}$ m$^3$.
This is equivalent to a ridge of 5 km height and total width at the
base of 10 km (assuming a nearly triangular section).

This multitude of collisions will lead necessarily to the sublimation
of at least some of the volatile components of these ring particles.
The conversion of a small percentage of the collisional energy to
thermal energy would be enough to totally sublimate some of the ices commonly
found at 10 a.u. from the Sun, like CO$_2$. The sublimation of such
volatiles not only act as as a powerful coolant of the whole process,
but it would also produce a transient atmosphere at the location of
the impacts, with a pressure decreasing symmetrically with distance
from the impact zone. The thin but fast winds resulting from this
pressure gradient could have carried large amounts of the refractory
ring particle ``dust'' away from the area where the ring material is
being deposited; a similar phenomenon is observed on comets; where ice
sublimation (due in that case to solar irradiation) carries dust into
space, producing the cometary dust tails. However, Iapetus has a
gravitational field that is much larger than the common comet,
therefore much of this dust does not escape into space, being instead
deposited around the impact zone. In our hypothesis, this is the dark
coating of the region known today as Cassini Regio.

Because of its unique nature, we will henceforth refer to the
equatorial ridge of Iapetus simply as ``the rindge'', to mean that
this feature is not a ridge in the usual sense of the term;
i.e., a mountain chain caused by tectonic processes.

\section{Tests.}

The dark streaks observed at the edge of Cassini Regio indicate that it
was a wind blowing from the equator that deposited the ``dust'', as
one would expect from our hypothesis. We can be certain of this because
the Cassini imagery shows clearly that the dust is deposited downwind
from crater rims. The same phenomenon is
observed near Martian craters after dust storms. Porco~et~al. (2005)
exclude this scenario because of the lack of an atmosphere on Iapetus,
and conclude that the particles flew in ballistic trajectories from the
equator, but this process can not deposit dust particles
preferably downwind from crater rims. Ballistic flight might, however,
have happened for larger, boulder-sized particles nearer the rindge.

A key prediction of this scenario is that the coating of ``dust''
should become progressively thinner away from the rindge: deposition of
dust by a fluid should be proportional to its concentration in that
fluid. If the deposition is significant, it will cause the dust
concentration in the fluid to slowly diminish away from the equator,
leading to smaller deposition per unit area with increasing latitude.
This is quite apparent in the albedo and color evolution
with latitude (Porco et al. 2005), such a pattern should not be
observed for ballistic flight. Another, still untested prediction
is that the dust particle size should be larger closer to the rindge;
because fluids tend to deposit larger particles first. This has not
yet been confirmed.

It has been suggested Porco et al. (2005) that the rindge could have
been created away from the equator and them moved towards it by energy
loss from the interior (Peale, 1977). However, the structure does not
show the complex tectonic patterns produced by such migration
(Perchman and Melosh, 1979). The ring collision scenario
naturally produces a linear feature exactly at the equator: this is
the geometric intersection of a ring plane and the surface of a moon
with a (previously) equatorial orbit (more on this later). Tectonic
processes are not likely to produce such a perfectly linear feature.
Tectonism is unlikely on Iapetus in any case; there are no clear
volcanic landforms nor is there a strong internal heat source; and the
object does not seem to be differentiated.
 
Another key feature of the rindge is that its height varies extremely
slowly with longitude. This is to be expected from a collision with
a ring, but such a constant height has never been observed for any
tectonic feature.

If the origin of the rindge was tectonic and preceded the dark coating,
then it should not necessarily be confined to Cassini Regio. If
it postdated the coating, then the rindge, being built from an
upwelling from the interior of Iapetus should be much brighter than
the surrounding surface. The ring collision scenario predicts that the
rindge should be confined to Cassini Regio (i.e. the rindge, being an
accumulation of collisional debris, must necessarily be surrounded by
the resulting ejecta) and have the same albedo, as observed.

\section{An impact with a ring edge.}

What was exactly the geometry of the collision of Iapetus with the
ring? Fortunately, the imaging provides us with abundant clues.
The longitudinal extent of the rindge is about 110$^\circ$ (Porco et
al. 2005). This
indicates that the Iapetus was never fully inside the ring region,
otherwise the longitudinal extent of the rindge would be at least
180$^\circ$. Simple considerations of orbital mechanics indicate that
a collision of a satellite with a ring edge should always cause an
eastwards motion of the particles relative to most of the satellite's
surface exposed to the impacts (see Figure 1).

This accounts for an important observed fact: although Cassini Regio
is symmetrical relative to the rindge in the North-South direction, it is
not so in the East-West direction. In particular, in a collision with
a ring edge the rindge should be taller on its Western side, as the
impacts there were closer to vertical and therefore more numerous per
unit area (see Figure 1), and should slowly diminish in height towards
the east, becoming imperceptible, as observed. These considerations
should be enough to make a numerical model capable of predicting the
longitudinal height profile of the rindge. Such a model will be
published elsewhere.

\section{Sudden orbital migration of Iapetus.}

The existence of the rindge suggests that the former orbit of Iapetus
was equatorial, otherwise, with its present inclination, a collision
with a ring would not produce a sharp rindge, but something more like a
wispy dark coating of the leading hemisphere\footnote{This is a potential
explanation to the features observed in the leading hemispheres of
Rhea and Dione.}. Because rings are generally located much closer to
Saturn than the present orbit of Iapetus (generally within a couple of
planetary radii of the surface, where tidal effects prevent
coalescence of the ring particles into larger objects, a region
known as the ``Roche zone''), a collision with a ring also suggests
that Iapetus was once much closer to Saturn.

The fact that the collision has occurred indicates a sudden
change of its orbital parameters prior to the ring collision, in
particular a large increase in the eccentricity (but not the
inclination) of the orbit, otherwise the
ring would have time to adjust to the gravity of Iapetus and no
collisions would occur (as observed for the satellites embedded in the
rings). The cause of this sudden change of orbit could be an
interaction with other satellites of Saturn or even an intruding
object, but the fact that the orbit of Iapetus was equatorial during
the ring collision strongly favors an interaction with another satellite
of Saturn. A sudden change of eccentricity would generally be
associated with a sudden change of orbital period. That would
make the rotation of the satellite asynchronous for some time, so the
leading hemisphere at the time of the impact is not necessarily the
leading hemisphere observed today; the model indicated above (see also
Fig. 1) suggests that the center of the leading hemisphere at the time
of the collision was closer to the western edge of the rindge, which
is now nearer the anti-Saturn point.

The sudden increase of eccentricity implied by the collision with a
ring is likely to lead to further orbital changes: satellites in
eccentric orbits and low inclinations have much increased
probability of crossing the orbits of other satellites and therefore
interact gravitationally with them.

The present orbit of Iapetus is an enigma. It is the most inclined of
all the regular satellites of Saturn (about 7$^\circ$). This orbit is
stable, i.e., Iapetus never comes close to Titan. However, this
implies that no interaction with extant satellites could have made
Iapetus change from an eccentric orbit with small perisaturnium to its
present orbit, which has a perisaturnium more than twice as large as
the aposaturnium of Titan. The transition from a previously equatorial
orbit closer to Saturn to its present orbit is the major conceptual
problem with the scenario discussed above. Its is likely to be a
low-probability event, probably caused by an interaction with a now
lost object in the general vicinity of the present orbit of Iapetus.

The low probability of such an event would have profound consequences.
Many proto-satellites formed in the Saturnian system were in
orbits that started, or later became, unstable. In the later stages of
satellite formation, proto-satellites in such orbits either end up
colliding with other proto-satellites, or are ejected from the Saturnian
system by close encounters with other proto-satellites, or get tidally
disrupted by close approaches to Saturn (forming a ring system). Many
objects have to go through such processes to account for the formation
of the large satellites, in particular Titan. This scenario, a small
replica of what is thought to have happened with the early solar
system (Cassen and Woolum, 1999), is the ``canonical'' explanation for
the formation of the large satellites of the giant planets (Buratti,
1999). With so many proto-satellites in chaotic orbits, collisions
with rings are a definite possibility for many objects. If the
survival of one of them in a distant orbit like that of Iapetus is a
small probability event, then the mere existence of Iapetus would imply
a large number of such primordial objects. That would add support for
the current understanding of planetary (and large satellite) formation.

%

\begin{acknowledgments}
The author would like to thank discussions and encouragement from the
Arecibo Observatory staff, in particular Michael Nolan, Steve
Torchinsky and Avinash Deshpande.
\end{acknowledgments}

%
%
%
%
%
%
%
%


%
%
%


\begin{figure}
\setlength{\unitlength}{1in}
\begin{picture}(0,1.8)
\put(0.0,0.2){\includegraphics{./fig1.eps}}
\end{picture}
 \caption{A collision with a ring edge should always
produce an eastward motion of the ring particles relative to the
surface of the satellite being exposed. {\it top} In one scenario,
Iapetus only meets the ring are the aposaturnium of its transient
orbit. In this case, the ring particles are moving faster towards the
east than Iapetus, resulting in eastwards motion relative to the
surface and a taller Western end of the rindge.
{\it bottom} In the alternative scenario, Iapetus touches the ring at
the perisaturnium of its transient orbit, moving faster towards the
East than the ring particles. Again, an eastwards motion of the
particles relative to the exposed surface is observed. We believe that
this is what eventually occurred, because the evidence points towards
a single ring collision and Iapetus is now much further out than the
outer rings of Saturn. The thick arc
indicates the forming rindge, the letter ``c'' indicates where the
``dust'' is being deposited. This is the region now known as Cassini
Regio.}
\end{figure}

%
%

\end{article}

\end{document}